\documentclass[amsmath, amsfonts, superscriptaddress, showpacs, twocolumn, prl]{revtex4}
\usepackage{graphicx}
\usepackage{epsfig}
\usepackage{bm}
\usepackage{dcolumn}
\usepackage{amsmath}
\usepackage{amssymb}
\usepackage{color}

\begin{document}

\title{Quantum-fluctuation effects in transport properties of superconductors
\\ above the paramagnetic limit}

\author{M.~Khodas}
\affiliation{Department of Physics and Astronomy, University of
Iowa, Iowa City, IA 52242, USA}

\author{A.~Levchenko}
\affiliation{Department of Physics and Astronomy, Michigan State
University, East Lansing, Michigan 48824, USA}

\author{G.~Catelani}
\affiliation{Departments of Physics and Applied Physics, Yale
University, New Haven, CT 06520, USA}

\begin{abstract}
We study the transport in ultrathin disordered film near the quantum
critical point induced by the Zeeman field. We calculate corrections
to the normal state conductivity due to quantum pairing
fluctuations. The fluctuation-induced transport is mediated by
virtual rather than real quasi-particles. We find that at zero
temperature, where the corrections come from purely quantum
fluctuations, the Aslamazov-Larkin paraconductivity term, the
Maki-Thompson interference contribution and the density of states
effects are all of the same order. The total correction leads to the
negative magnetoresistance. This result is in qualitative agreement
with the recent transport observations in the parallel magnetic
field of the homogeneously disordered amorphous films and
superconducting two-dimensional electron gas realized at the oxide
interfaces.
\end{abstract}

\date{February 22, 2012}


\maketitle

\textit{Introduction}.-- According to the microscopic
BCS-theory~\cite{BCS} magnetic field extinguishes superconductivity.
In the absence of spin-orbit interaction there are two basic
mechanisms. The first one is diamagnetic effect associated with the
action of the magnetic field on the orbital motion of electrons
forming a Cooper pair. The second, paramagnetic mechanism, is due to
Zeeman splitting of the states with the same spatial wave function
but opposite spin directions. In the former case, the estimate for
the upper critical field follows from the condition
$H_{c2}\xi^2\simeq\Phi_0$, where $\Phi_0=hc/2e$ is the flux quantum,
$\xi=\sqrt{\hbar D/\Delta}$ is the coherence length for the
disordered superconductor with $\Delta$ being energy gap, and $D$
diffusion coefficient. In contrast, Zeeman splitting destroys
superconductivity at the other critical field that follows from the
condition $g_L\mu_BH_z\simeq\Delta$, where $\mu_B=e\hbar/2mc$ is the
Bohr magneton and $g_L$ is renormalized giro-factor. The ratio
between two fields is $H_{z}/H_{c2}\sim k_F\ell\gg1$, where $k_F$ is
Fermi momentum and $\ell$ is the elastic scattering length. Thus, in
bulk systems, the suppression of superconductivity is typically
governed by the first -- diamagnetic mechanism. The situation
changes in the case of restricted dimensionality. For example, in
the case of thin-film superconductor the above ratio changes to
$H_z/H_{c2}\sim (k_F\ell)(d/\xi)$, which can be small provided that
film is thin enough $d\ll\xi/k_F\ell$, such that spin effects
dominate.

\begin{figure}[t!]
\includegraphics[width=8cm]{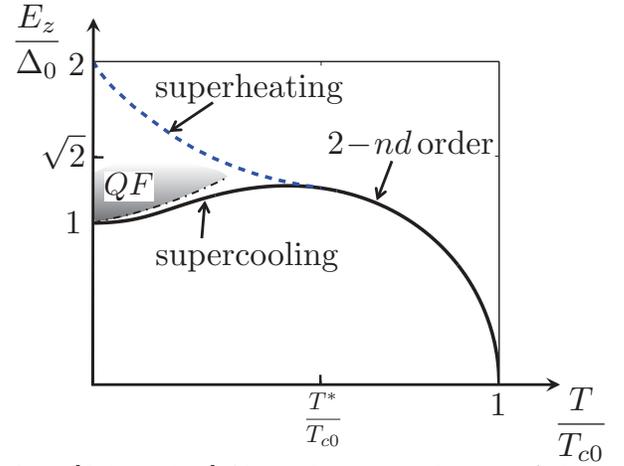}\vskip-.5cm
  \caption{[Color online] Above the tricritical point $T^*$ the
  second order paramagnet to superconductor transition occurs along the
  (black) solid line obtained from Eq.~\eqref{prop}.
  At $T<T^*$ this line becomes a supercooling part of the hysteresis,
  and the (blue) dashed line is its superheating part.
  The latter is obtained following Ref.~\cite{Suzuki-JPSJ}.
  The grey shaded area with the critical point $(0,\Delta_0)$
  as its lowest corner bounded by the black dashed line
  marks the region of quantum fluctuations.}\label{Fig1}
   \vskip-.5cm
\end{figure}

The scenario of paramagnetically limited superconductivity has long
history that goes back to pioneering works of Clogston and
Chandrasekhar~\cite{CC}. The first order phase transition from
superconductor to paramagnet was found at the critical field
approaching $E_z=\sqrt{2}\Delta$ at low temperatures. In practice,
the measured film resistance follows a hysteresis loop~
\cite{Wu:1994,Butko-PRL99,Wu:2006} instead of a sharp first order
transition. At low field $E_z<\sqrt{2}\Delta$ and zero temperature
the system is superconducting. With increasing the field $
\sqrt{2}\Delta<E_z<2\Delta$ the film is trapped in a superconducting
metastable state. At fields exceeding the superheating threshold
$E_z>2\Delta$ the film becomes normal. When the field is reduced
back to zero, the normal state is metastable in the interval
$\Delta<E_z<\sqrt{2}\Delta$~\cite{FFLO}. In this paper we study the
transport properties at the onset of transition to the
superconductivity near the supercooling field $E^{sc}_z = \Delta$
[see Fig.~\ref{Fig1}]. This field $E^{sc}_z(T)$ corresponds to the
zero binding energy of a Cooper pair and can be determined from the
standard equation~\cite{Fulde}
\begin{equation}\label{prop}
\ln(T_c/T_{c0})=\psi(1/2)-\mathrm{Re}\,\psi(1/2+i E^{sc}_z/4\pi T_c)
\end{equation}
similar to that in the theory of paramagnetic impurities~\cite{AG}.
Here $\psi$ is the digamma function and $T_{c0}=T_{c}(H=0)$ is the
critical temperature in the absence of a magnetic field. The zero
temperature solution of Eq.~\eqref{prop}, $E^{sc}_z(0) = \Delta$,
defines the quantum critical point (QCP), which is premier interest
of our study.

\textit{Motivation}.-- The renewed interest in the physics of
paramagnetically limited superconductors is motivated by the rapid
growth of its experimental realizations. Recent parallel magnetic
field studies of two-dimensional superconducting systems were
extended to much lower temperatures thus making it feasible to
approach the limit of QCP. Tunneling spectroscopy of ultrathin Al
and Be films revealed field-induced spin mixing and anomalous
resonances in the density of
states~\cite{Butko-PRL99,Adams-PRL04,Catelani-PRB09}. The latter was
successfully explained in
theory~\cite{Aleiner-PRL97,Catelani-PRB06}, which emphasized the
crucial role of superconducting pairing correlations in the
paramagnetic state even far from the transition region. A surprising
enhancement of superconductivity by a parallel magnetic field,
deduced from the transport measurements, was observed in ultrathin,
homogeneously disordered amorphous Pb films and the two-dimensional
electron gas realized at the interface of oxide insulators LaAlO$_3$
and SrTiO$_3$~\cite{Gardner-NatPhys11}. In addition, pronounced
negative magnetoresistance (NMR), concomitant with the enhanced
$T_c$, was reported. Although we do not dwell onto the issue of
$T_c$ enhancement in these systems (see Ref.~\cite{Michaeli} for the
recent theoretical proposals), we show that transport anomalies,
such as NMR, can be successfully addressed within BCS theory.

The issue of NMR in superconductors, either near the QCP or near the
parallel-field-tuned superconductor-insulator transition, was
previously discussed in the literature
experimentally~\cite{Gantmaher-JETP03,Parendo-PRB04} and attributed
theoretically~\cite{Galitski-PRB01,Lopatin-PRL05,Glatz-PRB11} to the
proliferation of superconductive fluctuations~\cite{Varlamov-Book}.
These studies emphasized mainly the orbital effect of a magnetic
field on the pre-formed Cooper pairs. In this work we develop
transport theory of paramagnetically limited ultrathin
superconductors focusing on the quantum regime of zero temperature
near the critical Zeeman field. The regime of classical fluctuations
was partially discussed in the early
papers~\cite{Fulde-Maki,Aoi,Tedrow,AHL}.

\textit{Theory}.-- In the vicinity of the transition transport
properties of superconductors are governed by the fluctuation
effects. These are famous paraconductivity phenomena introduced by
Aslamazov and Larkin (AL)~\cite{AL}, Maki and Thompson
(MT)~\cite{MT}, and also related density of states (DOS) effects
discussed first by Abrahams \textit{et al.}~\cite{DOS}. We follow
these classical papers and approach the problem based on the
diagrammatic perturbation theory. Note that the technique based on
the time-dependent Ginzburg-Landau formalism applied for studying
transport near QCP~\cite{Coleman-PRL97,Mineev-PRB01} accounts
correctly only for the classical part of AL-type contribution to the
conductivity, but it misses completely the quantum zero-temperature
corrections. Microscopic approach takes care of all the
contributions including DOS part, resulting from the depletion of
the normal state density of states by superconducting fluctuations,
and also MT interference
term~\cite{Galitski-PRB01,Lopatin-PRL05,Glatz-PRB11}. In fact, at
$T=0$ where the corrections come from purely quantum fluctuations,
these effects turn out to be of the dominant nature. In calculations
we assume diffusive limit,
\begin{equation}\label{Diff-Limit}
T\ll E_z,\Delta \ll\tau^{-1}\ll\varepsilon_F\, .
\end{equation}
Conditions \eqref{Diff-Limit} are satisfied in many experiments
\cite{Butko-PRL99,Adams-PRL04}.

Within Kubo linear response formalism conductivity is obtained from
$\sigma=-K^R(\omega)/i\omega$ by analytic continuation of the
Matsubara current correlation kernel
$K(\omega_n)=-\int^{1/T}_{0}d\tau e^{i\omega_n\tau}\langle T_\tau
J(\tau)J(0)\rangle$. This kernel can be conveniently presented as a
sum of three contributions $K=K_{AL}+K_{MT}+K_{DOS}$. The general
expression for the AL term reads (hereafter $\hbar=k_B=1$):
\begin{equation}\label{K-AL}
K_{AL}(\omega_n)=-e^2T\sum_{Q,\Omega_k}B^2_{Q,\Omega_k,\omega_n}
L_{Q,\Omega_k}L_{Q,\Omega_k+\omega_n},
\end{equation}
where $\Omega_k=2\pi kT$. The triangular vertex function
\begin{eqnarray}\label{B}
B_{Q,\Omega_k,\omega_n}\!\!=T\!\!\sum_{\sigma,\varepsilon_m}\!
\lambda^\sigma_{Q,\varepsilon_{m+n},\Omega_k-\varepsilon_m}
\lambda^\sigma_{Q,\varepsilon_n,\Omega_k-\varepsilon_n}J_{AL}^{\sigma},\\
J_{AL}^{\sigma}=\sum_P
v_PG^\sigma_{P,\varepsilon_{n+m}}G^\sigma_{P,\varepsilon_n}
G^{-\sigma}_{-P+Q,-\varepsilon_n+\Omega_k},
\end{eqnarray}
consists of two Cooperons
\begin{equation}\label{Cooperon}
\lambda^\sigma_{Q,\varepsilon_n,\varepsilon_m}=
\frac{\theta(-\varepsilon_n\varepsilon_m)}{\tau(DQ^2+|\varepsilon_n-\varepsilon_m|-i\sigma
E_z\mathrm{sgn}(\varepsilon_n-\varepsilon_m)}
\end{equation}
and an integral over the block of three Green's functions with
$G^\sigma_{P,\varepsilon_n}=(i\varepsilon_n-\xi_P+\sigma
E_z/2+\mathrm{sgn}(\varepsilon_n)/2\tau)^{-1}$. Here we used
notations: $\varepsilon_m=2\pi T(m+1/2)$,
$\xi_P=P^2/2m-\varepsilon_F$, $v_P=\partial_P\xi_P$,
$\theta(\varepsilon)$-step function and
$\mathrm{sgn}(\varepsilon)$-sign function. Finally, propagator of
fluctuating Cooper pairs in Eq.~\eqref{K-AL} is given by
\begin{equation}\label{L}
L^{-1}_{Q,\Omega_k}\!\!=-\nu\left[\ln\frac{T}{T_{c0}}-\psi\left(\frac{1}{2}\right)
+\frac{1}{2}\sum_{\sigma=\pm}\Psi^\sigma_{Q,\Omega_k}\right]
\end{equation}
where
$\Psi^\sigma_{Q,\Omega_k}=\psi\left(\frac{1}{2}+\frac{DQ^2+|\Omega_k|+i\sigma
E_z}{4\pi T}\right)$. When calculating $B$-vertex one should follow
few basic steps~\cite{Varlamov-Book}. \textit{i}) To the leading
order in the momentum transferred $Q$ one can approximate
$G^{-\sigma}_{-P+Q,-\varepsilon_n+\Omega_k}\approx
G^{-\sigma}_{P,-\varepsilon_n+\Omega_k}+(v_P\cdot
Q)(G^{\sigma}_{P,-\varepsilon_n+\Omega_k})^2$. \textit{ii})
Furthermore, one can neglect Zeeman energy as compared to the
inverse scattering time in the Green's functions [provided the
condition of Eq.~\eqref{Diff-Limit}] and then completes
$P$-integration in a standard way $\sum_P\to\nu\int
d\xi_P\int\frac{dO_P}{2\pi}$. \textit{iii}) Next is the fermionic
Matsubara $\varepsilon_m$-sum in Eq.~\eqref{B}, which can be found
in the closed form with the result
\begin{eqnarray}\label{B-AL}
B_{Q,\Omega_k,\omega_n}=\frac{\nu
Q_xD}{\omega_n}\sum_{\sigma}\left[\Psi^\sigma_{Q,|\Omega_k|+\omega_n}-
\Psi^\sigma_{Q,|\Omega_k|}\right.\nonumber\\
\left.+\Psi^\sigma_{Q,|\Omega_{k+n}|+\omega_n}-
\Psi^\sigma_{Q,|\Omega_{k+n}|}\right].
\end{eqnarray}
\textit{iv}) The remaining step of calculation is bosonic
$\Omega_k$-sum followed by an analytical continuation
$i\omega_n\to\omega$. The latter are accomplished via the contour
integration over the circle with two-brunch cuts at
$\mathrm{Im}\Omega=0,-\omega_n$ where the product of propagators in
Eq.~\eqref{K-AL} has breaks of analyticity. After $\omega$-expansion
of $K^R_{AL}(\omega)$ to the linear order one finds for the AL
conductivity correction
$\sigma^{AL}=\sigma^{AL}_{cl}+\sigma^{AL}_{q1}+\sigma^{AL}_{q2}$,
where
\begin{eqnarray}
&&\hskip-.25cm\sigma^{AL}_{cl}=\frac{e^2}{4\pi
T}\sum_Q\int^{+\infty}_{-\infty}\frac{d\Omega}{\sinh^2\frac{\Omega}{2T}}
(B^{RA}_{Q,\Omega})^2(\mathrm{Im}L^R_{Q,\Omega})^2,\\
&&\hskip-.25cm\sigma^{AL}_{q1}=\frac{e^2}{4\pi}\sum_Q\int^{\infty}_{0}
d\Omega\coth\frac{\Omega}{2T}\nonumber\\
&&\hskip-.25cm\times\mathrm{Re}\{[(B^{RA}_{Q,\Omega})^2-(B^{RR}_{Q,\Omega})^2]
\partial_\Omega(L^R_{Q,\Omega})^2\},\label{sigma-al-q1}\\
&&\hskip-.25cm\sigma^{AL}_{q2}=-\frac{e^2}{4\pi}\int^{+\infty}_{-\infty}
d\Omega\coth\frac{\Omega}{2T}\nonumber\\
&&\hskip-.25cm\times\{\partial_\omega(B^{RR}_{Q,\Omega,\omega})^2(L^R_{Q,\Omega})^2
-\partial_\omega(B^{AA}_{Q,\Omega-\omega,\omega})^2(L^A_{Q,\Omega})^2\nonumber\\
&&\hskip-.25cm+\partial_\omega[(B^{RA}_{Q,\Omega-\omega,\omega})^2-(B^{RA}_{Q,\Omega,\omega})^2]
|L^R_{Q,\Omega}|^2\}.\label{sigma-al-q2}
\end{eqnarray}
The superscripts $R/A$ in the vertex function and propagators stand
for the retarded/advanced components while subscripts $cl/q$ refer
to classical/quantum. This convention comes form the observation
that as $T\!\to\!0$ classical contribution vanishes while quantum
remains finite.

We turn now to the derivation of the MT contribution whose response
kernel is given by
\begin{equation}\label{K-MT}
K_{MT}(\omega_n)=e^2T\sum_{\Omega_k,Q}L_{Q,\Omega_k}\Sigma^{MT}_{Q,\Omega_k,\omega_n}
\end{equation}
where
\begin{eqnarray}\label{Sigma-MT}
&&\hskip-.8cm\Sigma^{MT}_{Q,\Omega_k,\omega_n}\!\!=T\!\!\sum_{\sigma,\varepsilon_m}\!
\lambda^\sigma_{Q,\varepsilon_{m+n},\Omega_{k-n}-\varepsilon_m}\!
\lambda^{\sigma \mathrm{sgn}[\epsilon_m \epsilon_{m+n}]}_{Q,\varepsilon_m,\Omega_k-\varepsilon_m}\!J_{MT}\\
&&\hskip-.8cm J_{MT}=\sum_P
v_Pv_{Q-P}G^\sigma_{P,\varepsilon_{m+n}}
G^{-\sigma}_{Q-P,\Omega_{k-n}-\varepsilon_m}\nonumber\\
&&\phantom{G^{-\sigma}_{Q-P,-\varepsilon_m-\omega_n+\Omega_k}}
\times
G^\sigma_{P,\varepsilon_m}G^{-\sigma}_{Q-P,-\varepsilon_m+\Omega_k}.
\end{eqnarray}
Momentum integration in the block of Green functions $J_{MT}$ is
done under the same approximations as in the case of AL term
described above. According to the standard
convention~\cite{Varlamov-Book} we split now MT term into the
so-called regular and anomalous contributions:
\begin{subequations}
\begin{equation}
\Sigma^{MT(reg)}_{Q,\Omega_k,\omega_n}\!=-\frac{\nu
D}{\omega_n}\sum_\sigma[\Psi^\sigma_{Q,|\Omega_k|+2\omega_n}-\Psi^{\sigma}_{Q,|\Omega_k|}],
\end{equation}
\begin{equation}
\Sigma^{MT(an)}_{Q,\Omega_k,\omega_n}\!=-\frac{\nu
D}{2(DQ^2\!+\!\omega_n)}\sum_{\sigma}
[\Psi^\sigma_{Q,-|\Omega_k|+2\omega_n}-\Psi^{\sigma}_{Q,|\Omega_k|}].
\end{equation}
\end{subequations}
After the analytical continuation these translate into the
conductivity correction
$\sigma^{MT}=\sigma^{MT}_{reg}+\sigma^{MT}_{an}$, where
\begin{subequations}
\begin{equation}\label{sigma-mt-reg}
\sigma^{MT}_{reg}=-\frac{e^2\nu D}{8\pi^3T^2}\sum_{\sigma
Q}\int^{\infty}_{0}d\Omega\coth\frac{\Omega}{2T}
\mathrm{Im}[L^R_{Q,\Omega}(\Psi^\sigma_{Q,-i\Omega})''],
\end{equation}
\begin{equation}
\sigma^{MT}_{an}=\frac{e^2\nu D}{8\pi T}\sum_{\sigma
Q}\int^{+\infty}_{-\infty}\!\!\frac{d\Omega}{\sinh^2\frac{\Omega}{2T}}
\frac{L^R_{Q,\Omega}[\Psi^\sigma_{Q,i\Omega}-\Psi^\sigma_{Q,-i\Omega}]}
{DQ^2+\Gamma_{\phi}}.
\end{equation}
\end{subequations}
In order to regularize logarithmically divergent momentum integral
in the case of anomalous contribution we have introduced
pair-breaking cutoff parameter $\Gamma_{\phi}$.

We finally discuss the density of states contribution to the
conductivity. The latter is given by the similar to Eq.~\eqref{K-MT}
expression with
\begin{equation}\label{K-DOS}
K_{DOS}(\omega_n)=e^2T\sum_{\Omega_k,Q}L_{Q,\Omega_k}\Sigma^{DOS}_{Q,\Omega_k,\omega_n}
\end{equation}
where
\begin{eqnarray}\label{Sigma-DOS}
\Sigma^{DOS}_{Q,\Omega_k,\omega_n}=2T\sum_{\sigma,\varepsilon_m}
(\lambda^\sigma_{Q,\varepsilon_m,\Omega_k-\varepsilon_m})^2J_{DOS},\\
J_{DOS}=\sum_{P}v^2_P(G^\sigma_{P,\varepsilon_m})^2G^\sigma_{P,\varepsilon_m+\omega_n}
[G^{-\sigma}_{Q-P,\Omega_k-\varepsilon_m}\nonumber\\+\frac{1}{2\pi\nu\tau}\sum_{P'}
(G^\sigma_{P',\varepsilon_m})^2G^{-\sigma}_{Q-P',\Omega_k-\varepsilon_m}].
\end{eqnarray}
After standard steps outlined above one arrives at the
conductivity correction
$\sigma^{DOS}=\sigma^{DOS}_{cl}+\sigma^{DOS}_{q}$ in the form
\begin{subequations}
\begin{equation}\label{sigma-dos-cl}
\sigma^{DOS}_{cl}\!=\!-\frac{e^2\nu D}{16\pi^2T^2}\!\sum_{\sigma
Q}\!\!\int^{+\infty}_{-\infty}\!
\frac{d\Omega[(\Psi^\sigma_{Q,i\Omega})'-(\Psi^\sigma_{Q,-i\Omega})']
}{\sinh^2\frac{\Omega}{2T}}L^R_{Q,\Omega},
\end{equation}
\begin{equation}\label{sigma-dos-q}
\sigma^{DOS}_{q}=  \sigma^{MT}_{reg}.
\end{equation}
\end{subequations}
The equality between the two contribution in Eq.~\eqref{sigma-dos-q}
has parallels with the original fluctuation transport considerations
at $T-T_c\ll T$. In the original near--$T_c$ problem, the typical
energy of diffusing pairs $D Q^2\sim T-T_c$  is smaller than the
thermal energy of quasiparticle $\sim T$. In our case, $E_z$ adds to
the energy of pairs making it bigger than $T$. Correspondingly,
unlike the near--$T_c$ case, the off--shell energy of a pair,
$2\varepsilon \sim T$, falls below the pair excitation energy set by
$E_z$. This causes a sign inversion of the energy denominator
associated with the unbound intermediate state and the correction
\eqref{sigma-dos-q} turns to be positive. In general, derived above
conductivity corrections are applicable at any field $H$ and
temperature $T$ above the transition. In the following we discuss
limiting case of interest.

\textit{Results}.-- It is convenient to regroup all contributions
and present total conductivity correction as the a sum of
zero-temperature $(\delta\sigma_q)$ and finite-temperature
$(\delta\sigma_T)$ terms, namely
\begin{equation}
\delta\sigma(H,T)=\delta\sigma_q(H)+\delta\sigma_T(H,T).
\end{equation}
The first term here is determined by the quantum AL
[Eqs.~\eqref{sigma-al-q1}-\eqref{sigma-al-q2}] and DOS
[Eq.~\eqref{sigma-dos-q}] contributions, and also regular part of
the MT conductivity [Eq.~\eqref{sigma-mt-reg}]. The remaining terms
define $\delta\sigma_{T}$. The magnitude of $\delta\sigma_q$
decreases monotonically with increasing field; this leads to a
\textit{negative} magnetoresistance at zero temperature. At finite
temperature, based on how the quantum critical point is approached,
there are several regimes that show different $T$ and $H$
dependencies, which should be experimentally accessible. Below we
focus on QCP only and extract the leading singularity in
$\delta\sigma_q$ as the function of Zeeman field. Thermal
contribution $\delta\sigma_T$ and various crossover regimes will be
discussed elsewhere~\cite{KLC}.

At zero temperature $\Psi^{\sigma}_{Q,\pm i\Omega}\to\ln[(DQ^2\pm
i\Omega+i\sigma E_z)/4\pi T]$ and the pair-propagator can be taken
in the leading pole approximation
\begin{equation}
L^{R(A)}_{Q,\Omega}\approx-\frac{2\Delta^2_0/\nu}{E^2_c-(\Omega\pm
iDQ^2)^2},
\end{equation}
which is obtained from Eq.~\eqref{L} under the conditions
$DQ^2\ll\Delta_0$ and $|E_c\pm\Omega|\ll\Delta_0$. Here
$\Delta_0=\pi T_{c0}/2\gamma_E$ where $\ln\gamma_E\approx0.57$ is
the Euler constant, and $E_c=\sqrt{E^2_z-\Delta^2_0}$. The branch
cut of the propagator (due to the logarithmic structure) also
contributes to $\delta\sigma_q$ but gives the sub-leading
singularity. Within the same accuracy we compute vertex functions:
\begin{equation}\label{VF1}
(B^{AA(RR)}_{Q,\Omega,\omega})^2=\frac{8\nu^2D}{E^4_z}DQ^2(DQ^2\pm
i\Omega)(DQ^2\pm i\Omega-2i\omega),
\end{equation}
\begin{equation}\label{VF2}
(B^{RA}_{Q,\Omega,\omega})^2=\frac{8\nu^2D}{E^4_z}(DQ^2)^2(DQ^2-2i\omega).
\end{equation}
All together this leads to the conductivity correction near the
Zeeman field-induced quantum critical point
\begin{equation}\label{sigma-result}
\delta\sigma_q(H) = \frac{2 e^2}{\pi^2 }
\ln\left(\frac{E_z}{E_z-\Delta_0}\right)
\end{equation}
which is obtained within the logarithmic accuracy. Equation
\eqref{sigma-result} is the main result of the paper.

\textit{Discussions}.-- The conceptual difference of our analysis
from the problem of fluctuation-induced transport close to $T_c$ is
that unpaired particles, have finite excitation energy $E_z$, see
Eq.~\eqref{Cooperon}. As a result, the activation probability of
such pairs is  suppressed exponentially $\propto \exp(-E_z/T)$. We
argue that while in the standard case the real gapless pairs are
only important in our case such pairs are always virtual.

Let us illustrate this point taking AL correction as an example.
Consider first standard case near--$T_c$. In Eq.~\eqref{K-AL} the
triangular vertex Eq.~\eqref{B-AL} can be estimated as
$B_{Q,\Omega,\omega} \propto D Q_x \partial \Pi_{Q,\Omega} /
\partial \Omega$. Here $\Pi_{Q,\Omega} = L^{-1}_{Q,\Omega} + g^{-1}$
is a particle-particle polarization operator with momentum $Q$
entering in a $D Q^2 - i \Omega$ combination. At small momenta we
can take $\Pi_{0,\Omega} $ in the clean system. The imaginary part
of the polarization operator $\mathrm{Im} \Pi \approx \int d \xi[n(-\xi_p +
\Omega)n( \xi_p )-\tilde{n}( -\xi_p + \Omega) \tilde{n}( \xi_p ) ]
\delta( \Omega - 2 \xi ) = \nu(\Omega/2) \tanh \frac{ \Omega}{ 2 T
}$, where the particle and hole occupation numbers are
$n(\varepsilon) =(1+e^{\varepsilon/T})^{-1}$,
$\tilde{n}(\varepsilon)=1-n(\varepsilon)$. The real part, due to
virtual pairs $\mathrm{Re} \Pi \approx \log \left| (\Omega^2 - T^2) /
\omega_d^2 \right| $, is a familiar Cooper logarithm.
The imaginary part contribution $B_{Q,\Omega,\omega}$
$ \propto DQ_x / T $.
In contrast, the real part contribution vanishes at $\Omega =0$ due to
the particle-hole symmetry, $\nu(\Omega) = \nu$. The expansion in
$\Omega \sim T - T_c \ll T$ yields a correction small in the
parameter $(T-T_c)/T_c \ll1$.

In the presence of Zeeman field the situation is very different. The
pair activation rate, $\mathrm{Im} \Pi \approx \nu(\Omega) [
n(\omega/2 - E_z/2) - n (\omega/2 + E_z/2)] $, gives exponentially
suppressed contribution $ \propto DQ_x \exp(-E_z/T)/ T $. The real
part, due to virtual pair excitation, can be obtained by the
Kramers-Kronig relation, $\mathrm{Re} \Pi  \approx  \log
\left|(\Omega^2 - E^2_z)/ \omega_d^2 \right| $. Its contribution to
$B_{Q,\Omega,\omega}$ is suppressed only algebraically $ \propto
DQ_x T / E_z^2 $. Unlike the standard case the virtual
quasi-particles make a dominant contribution to the triangular
vertex excitations. The algebraic suppression of vertexes is most
pronounced in the case of the AL and is manifested in additional
factors of $DQ^2$, $\Omega$ in Eq.~\eqref{VF1}-\eqref{VF2}, which
makes it logarithmic in $E_z/E_c$. Note that in the case of
near--$H_{c2}$ problem \cite{Galitski-PRB01} the AL contribution is
also suppressed due to the current matrix elements connecting
adjacent Landau levels

The regular MT and DOS contributions are proportional to a second
derivative of the real part of the polarization operator
$\mathrm{Re} \Pi_{\Omega,Q}$. Since the latter is finite at $\Omega
= 0$, these contributions are as singular as AL terms.

We have checked explicitly that other contributions
such as diffusion coefficient renormalization as well as
contribution with only one or no Cooperon vertexes are either small
or non-singular. Since the temperature can be set to zero in
integrations over fast fermion degrees of freedom, the additional
factors of $\tau$ results in small prefactors $\tau E_z$, $\tau D
Q^2 $ or $\tau \Omega$.

\textit{Outlook}.-- The spin-orbit scattering and finite thickness
effects modify the fluctuation transport, due to the finite spectral
weight in the particle-particle channel at zero frequency. Addition
of a finite spin-orbit scattering introduces a finite life time
$\Gamma^{-1}$ to the Cooperon. At lowest temperatures the
superconductivity survives if this scattering is not too strong,
$\Gamma \ll E_z$ with somewhat lower critical field. While $E_z$
approaches the supercooling transition from above the results
obtained in the present paper are expected to cross over to a
different regime at $\Gamma \approx E_c$. The finite film thickness
affects the crossover in a similar way. All these relevant
perturbations as well as the regime of close proximity to the
supercooling line will be studied elsewhere~\cite{KLC}.

We thank A.~Kamenev for very useful discussions and remarks. We also
thank P.~W.~Adams for the correspondence regarding the ongoing
transport experiments in ultrathin superconducting films. This work
was supported by University of Iowa (M.~K.), Michigan State
University (A.~L.) and Yale University (G.~C.).


\begin{thebibliography}{00}

\bibitem{BCS}
J.~Bardeen, L.~N.~Cooper, and J.~R.~Schrieffer, Phys. Rev.
\textbf{108}, 1175 (1957).

\bibitem{CC}
A.~M.~Clogston, Phys. Rev. Lett. \textbf{9}, 266 (1962);
B.~S.~Chandrasekhar, Appl. Phys. Lett. \textbf{1}, 7 (1962).

\bibitem{Wu:1994}
X.~S.~Wu, P.~W.~Adams, Phys. Rev. Lett. \textbf{73}, 1412 (1994).

\bibitem{Butko-PRL99}
V.~Yu.~Butko, P.~W.~Adams, and I.~L.~Aleiner, Phys. Rev. Lett.
\textbf{82}, 4284 (1999).

\bibitem{Wu:2006}
X.~S.~Wu, P.~W.~Adams, and G.~Catelani, Phys. Rev. B. \textbf{74}, 144519 (2006).


\bibitem{FFLO}
P.~Fulde and R.~A.~Ferrel, Phys. Rev. \textbf{135}, 550 (1964);
A.~I.~Larkin and Yu.~N.~Ovchinnikov, Sov. Phys. JETP \textbf{20},
762 (1965). Note that these papers also predict spatially
inhomogeneous state for $\sqrt{2}<E_z/\Delta<1.52$. We neglect such
possibility in this work.

\bibitem{Suzuki-JPSJ}
T.~Suzuki, Y.~Seguchi and T.~Tsuboi, J. Phys. Soc. Jpn. \textbf{69},
1462 (2000).

\bibitem{Fulde}
P.~Fulde, Adv. Phys \textbf{22}, 667 (1973).

\bibitem{AG}
A.~A.~Abrikosov and L.~P.~Gor'kov, Sov. Phys. JETP \textbf{12}, 1243
(1961).

\bibitem{Adams-PRL04}
P.~W.~Adams, Phys. Rev. Lett. \textbf{92}, 067003 (2004).

\bibitem{Catelani-PRB09}
G.~Catelani \textit{et al.}, Phys. Rev. B \textbf{80}, 054512
(2009).

\bibitem{Aleiner-PRL97}
I.~L.~Aleiner and B.~L.~Altshuler, Phys. Rev. Lett. \textbf{79},
4242 (1997); H.-Y.~Kee, I.~L.~Aleiner, and B.~L.~Altshuler, Phys.
Rev B \textbf{58}, 5757 (1998).

\bibitem{Catelani-PRB06}
G.~Catelani, Phys. Rev. B \textbf{73}, 020503(R) (2006).

\bibitem{Gardner-NatPhys11}
H.~J.~Gardner \textit{et al.}, Nat. Phys. \textbf{7}, 895 (2011).

\bibitem{Michaeli}
K.~Michaeli, A.~C.~Potter, P.~A.~Lee, preprint arXiv:1107.4352.

\bibitem{Gantmaher-JETP03}
V.~F.~Gantmaher \textit{et al.}, JETP Lett. \textbf{77}, 424 (2003);
JETP Lett. \textbf{71}, 473 (2000).

\bibitem{Parendo-PRB04}
K.~A.~Parendo \textit{et al.}, Phys. Rev. B \textbf{70}, 212510
(2004).

\bibitem{Galitski-PRB01}
V.~M.~Galitski and A.~I.~Larkin, Phys. Rev. B \textbf{63}, 174506
(2001).

\bibitem{Lopatin-PRL05}
A.~V.~Lopatin, N.~Shah, and V.~M.~Vinokur, Phys. Rev. Lett.
\textbf{94}, 037003 (2005).

\bibitem{Glatz-PRB11}
A.~Glatz, A.~A.~Varlamov, and V.~M.~Vinokur, Phys. Rev. B
\textbf{84}, 104510 (2011).

\bibitem{Varlamov-Book}
A.~I.~Larkin and A.~Varlamov, \textit{Theory of Fluctuations in
Superconductors} (Clarendon Press, Oxford, 2005).

\bibitem{Fulde-Maki}
P.~Fulde and K.~Maki, Z. Physik \textbf{238}, 233 (1970).

\bibitem{Aoi}
K.~Aoi, R.~Meservey and P.~M.~Tedrow, Phys. Rev. B \textbf{9}, 875
(1974).

\bibitem{Tedrow}
P.~M.~Tedrow and R.~Meservey, Phys. Rev. B \textbf{16}, 4825 (1977).

\bibitem{AHL}
A.~G.~Aronov, S.~Hikami, and A.~I.~Larkin, Phys. Rev. Lett.
\textbf{62}, 965 (1989).

\bibitem{AL}
L. G. Aslamazov and A. I. Larkin, Sov. Phys. Solid. State
\textbf{10}, 875 (1968).

\bibitem{MT}
K.~Maki, Prog. Theor. Phys. \textbf{39}, 897 (1968); R.~S.~Thompson,
Phys. Rev. B \textbf{1}, 327 (1970).

\bibitem{DOS}
E.~Abrahams, M.~Redi, and J.~W.~Woo, Phys. Rev. B \textbf{1}, 208
(1970).

\bibitem{Coleman-PRL97}
R.~Ramazashvili and P.~Coleman, Phys. Rev. Lett. \textbf{79}, 3752
(1997).

\bibitem{Mineev-PRB01}
V.~P.~Mineev and M.~Sigrist, Phys. Rev. B \textbf{63}, 172504
(2001).

\bibitem{KLC}
M.~Khodas, A.~Levchenko and G.~Catelani, unpublished.

\end{thebibliography}
\end{document}